\documentclass[a4paper,fleqn]{cas-dc}

\usepackage[numbers]{natbib}
\usepackage{bm}
\usepackage{graphicx}
\usepackage{caption}
\usepackage{subcaption}
\def\tsc#1{\csdef{#1}{\textsc{\lowercase{#1}}\xspace}}
\tsc{WGM}
\tsc{QE}
\tsc{EP}
\tsc{PMS}
\tsc{BEC}
\tsc{DE}

\begin{document}
\let\WriteBookmarks\relax
\def\floatpagepagefraction{1}
\def\textpagefraction{.001}

\shorttitle{FECT: Classification of Breast Cancer Pathological Images Based on Fusion Features}

\shortauthors{Hao et~al.}  

\title [mode = title]{FECT: Classification of Breast Cancer Pathological Images Based on Fusion Features}                      



%
\author[3]{Jiacheng Hao}[]
\fnmark[1]

\ead{haojiacheng1@gmail.com}



\affiliation[1]{organization={Institute of Biopharmaceutical and Health Engineering, Tsinghua Shenzhen International Graduate School },
    city={Shenzhen},
    postcode={518055}, 
    country={China}}

\author[1]{Yiqing Liu}[orcid=0000-0002-8157-2814]
\ead{liuyiqin20@mails.tsinghua.edu.cn}
\fnmark[1]
\author[2]{Siqi Zeng}[]
\ead{zengsq@tsinghua-gd.org}



\affiliation[2]{organization={Medical Optical Technology R\&D Center, Research Institute of Tsinghua, Pearl River Delta},
    city={Guangzhou},
    postcode={510700}, 
    country={China}}
\affiliation[3]{organization={School of Medicine, Nankai University},
    city={Tianjin},
    postcode={300071}, 
    country={China}}
\author%
[1]
{Yonghong He}
\cormark[1]
\ead{heyh@sz.tsinghua.edu.cn}


\cortext[cor1]{Corresponding author}

\renewcommand{\footnote}{\fnsymbol{13}}
\fntext[1]{These authors contributed equally to this work.}



\begin{abstract}
Breast cancer is one of the most common cancers among women globally, with early diagnosis and precise classification being crucial. With the advancement of deep learning and computer vision, the automatic classification of breast tissue pathological images has emerged as a research focus. Existing methods typically rely on singular cell or tissue features and lack design considerations for morphological characteristics of challenging-to-classify categories, resulting in suboptimal classification performance. To address these problems, we proposes a novel breast cancer tissue classification model that Fused features of Edges, Cells, and Tissues (FECT), employing the ResMTUNet and an attention-based aggregator to extract and aggregate these features. Extensive testing on the BRACS dataset demonstrates that our model surpasses current advanced methods in terms of classification accuracy and F1 scores. Moreover, due to its feature fusion that aligns with the diagnostic approach of pathologists, our model exhibits interpretability and holds promise for significant roles in future clinical applications.
\end{abstract}


\begin{highlights}
\item We propose a novel breast cancer tissue classification model that applies Fused features of Edges, Cells, and Tissues (FECT).
\item We design an edge feature extractor that uses an attention mechanism-based aggregator and a cell adjacency graph constructed with the KNN algorithm, to precisely differentiate between morphologically similar IC and DCIS.
\item We evaluate the performance of the FECT model on BRACS, the largest public breast cancer tissue classification dataset, with experiments demonstrating that our model significantly outperforms current state-of-the-art methods.
\end{highlights}


\maketitle

\section{Introduction}

Breast cancer is one of the most prevalent cancers among women worldwide. According to a report by the World Health Organization, breast cancer is the leading cause of cancer-related deaths among women \cite{bray2018global}. Early diagnosis and accurate pathological classification of this disease are crucial for improving treatment outcomes and survival rates for patients \cite{allemani2015global}. Traditionally, the diagnosis of breast cancer has relied on pathologists examining tissue slides under a microscope, a process that is not only time-consuming and labor-intensive but also limited by the individual experience and fatigue of the pathologists, leading to potential misdiagnoses and missed diagnoses. With the advent of digital pathology and advancements in computer vision, the use of computer-assisted pathology diagnostic systems to automatically identify and classify breast epithelial tissues has become a cutting-edge area of research \cite{bera2019artificial, 4niazi2019digital, 5campanella2019clinical,6kather2019deep,7lu2021data,8chen2024towards,9lu2024visual}. Deep learning analysis of Whole Slide Images (WSIs) can aid pathologists in faster and more consistent diagnoses, and is expected to play an increasingly critical role in future clinical settings.

There has been considerable research on the classification of breast tissue pathology images. \cite{10zhou2019cgc} utilized a graph convolutional network to construct a cell graph and applied graph clustering to effectively integrate cell morphology and microstructure information, enhancing the accuracy of cancer grading. \cite{11pati2022hierarchical} proposed a hierarchical graph neural network that sequentially processes the entity-graph representation to map tissue compositions to tissue subtypes. \cite{12patel2023garl}, a deep learning model combining a Graph Neural Network (GNN) with an adaptive regularization strategy was proposed to improve the accuracy of breast cancer image classification. However, GNNs may struggle with scalability in breast cancer classification due to the high computational demands of processing large and complex tissue graphs, potentially limiting their effectiveness in capturing the full spectrum of pathological features. \cite{13shao2021transmil} introduced a Transformer-based multiple instance learning method (TransMIL), which effectively addresses pathological image classification problems by exploring morphological and spatial information. \cite{14zhang2022attention} described a novel multiple instance learning approach (AMIL-Trans) that integrates channel attention and self-attention mechanisms for the classification of breast cancer whole slide images. \cite{15stegmuller2023scorenet} presented a Transformer-based architecture capable of efficiently identifying key areas in histopathological images, generating representations with generalizability. Despite the ability of the Transformer-based \cite{16dosovitskiy2020image} breast tissue classification model to effectively capture long-range dependencies in images, its integration of local and global features relies on effective design of data augmentation and attention mechanisms, which may perform poorly in recognizing specific pathological features without sufficient training data. \cite{17wang2024breast} presented a connectivity-aware graph transformer for breast cancer classification by phenotyping the topology connectivity of the tissue graph constructed from digital pathology images. Although this model achieved significant results in integrating local and global features and improving breast cancer classification accuracy, its performance might be influenced by the input graph structure; varying graph densities (e.g., sparser or denser graphs) could affect model performance, and determining the optimal graph structure might pose a challenge.

However, our empirical research has highlighted the challenges currently encountered in this task: 
\begin{enumerate}
\item {} \textbf{Dependence on a single feature, resulting in poor model performance and insufficient generalization.} Although existing models have introduced innovative deep learning image processing techniques and self-supervised learning methods into breast cancer classification, they typically rely on a single or limited set of features to perform the classification tasks. This approach may lead to performance degradation when dealing with highly heterogeneous breast pathology images, suggesting a potential benefit from fully utilizing the multi-level, multi-dimensional information present in breast cancer tissue images for integration. 
\item {} \textbf{A lack of design for morphological characteristics of difficult-to-classify categories.} For instance, benign pathologies or precancerous lesions may be confused with invasive carcinoma due to their atypical trabecular structures. In actual clinical diagnostic scenarios, pathologists need to make diagnoses based on the subtle differences in these fine structures. Therefore, it is necessary to consider incorporating this clinical prior knowledge into the model, to increase focus on these structures and enhance the model's interpretability.
\end{enumerate}

To address the issues identified, we propose a breast cancer tissue classification model that utilizes Fused features of Edges, Cells, and Tissues (FECT). In the cell feature extraction phase, we begin by accurately segmenting individual cells within the image, followed by training these single-cell features using an attention-based feature aggregator \cite{18vaswani2017attention} to obtain exhaustive cell features at the TROI (Tissue Region of Interest) level. Subsequently, we employ the high-performance ResMTUNet model \cite{19liu2024invasive} to extract tissue-level features. With precise segmentation labels on breast cancer tissue images, we utilize advanced image processing techniques to identify tissue edges and train specialized patch feature aggregators for these edge areas to extract critical edge features. Additionally, we apply the KNN algorithm to construct adjacency graphs for cell regions at tissue edges, capturing detailed micro-environmental distribution features of cells and tissues. Ultimately, by training with a Support Vector Machine (SVM) that combines features fused by weights, we have developed a model with outstanding classification performance in breast cancer tissue classification tasks. This not only enhances the model's accuracy but also significantly boosts its potential for application in complex clinical environments. In summary, our main contributions are as follows:

\begin{itemize} 
\item \textbf{We introduced a novel breast cancer tissue classification model, FECT, which incorporates fused features based on cells, tissues, and edges.} During the feature extraction process for cells and edges, the FECT model integrates an attention-based aggregator, effectively enhancing the representational capability of the features. This enhancement allows the FECT model to demonstrate higher accuracy in identifying and classifying complex types of breast cancer tissues. 
\item To precisely differentiate between two challenging types of cancer, invasive carcinoma and ductal carcinoma in situ, \textbf{we designed an edge feature extractor and employed the KNN algorithm to construct an adjacency graph of cells at the edges.} This approach meticulously captures the subtle differences and spatial relationships at the tissue edges, complementing pixel-level information with the spatial structure of the tumor microenvironment. This significantly improves the model's sensitivity and accuracy in distinguishing between these morphologically similar types of cancer. 
\item \textbf{We evaluated the performance of the proposed FECT on the largest publicly available breast cancer tissue classification dataset, BRACS\cite{20brancati2022bracs}.} Extensive empirical evidence supports the effectiveness of our approach, which significantly outperforms current state-of-the-art methods in terms of accuracy and sensitivity in classifying complex breast cancer tissue types.

\end{itemize}

\section{Related works}

\subsection{Tumor subtyping in digital pathology}

In the field of pathological image analysis, methods based on Convolutional Neural Networks (CNNs) such as VGG \cite{21simonyan2014very}, ResNet \cite{22he2016deep}, and Inception Network \cite{23szegedy2015going}, \cite{24chollet2017xception} have been widely applied for feature extraction and classification tasks in medical imaging \cite{25roy2019patch,26spanhol2016breast,27hameed2022multiclass,28vuong2021multi}. These traditional CNN models are effective in extracting local features from high-resolution images but are typically constrained by fixed receptive fields and limited ability to handle long-distance spatial dependencies. To overcome these limitations, the Vision Transformer model (ViT) \cite{16dosovitskiy2020image} has been introduced into pathological image analysis \cite{15stegmuller2023scorenet,29mehta2022end,30chen2022scaling}. ViT utilizes a self-attention mechanism, allowing for a more flexible handling of the relationships between different parts of an image. This makes Transformers particularly suitable for dealing with large-scale pathological images that feature complex structures and intercellular interactions, thereby offering deeper insights and improved classification performance.

Given that Whole Slide Images (WSIs) usually have high resolution and massive data volumes, and often lack detailed fine annotations of tumor locations, training deep neural networks directly on WSIs requires very large GPU memory, which is often unfeasible. Consequently, deep neural networks typically employ weakly supervised multiple instance learning (MIL) methods for histopathological analysis, analyzing pathological images in patches \cite{13shao2021transmil,31li2021dual,32lu2021data,33chen2023dmil,34tang2023multiple}. However, these methods primarily focus on analyzing pixel-level features such as texture, color, and shape. Recent studies highlight that breast cancer is a highly heterogeneous and complex tumor whose characteristics are influenced not only by the tumor itself but also significantly by the surrounding microenvironment, including adjacent blood vessels, immune cells, fibrous tissue, and the extracellular matrix. Considering these microenvironmental factors can provide a more in-depth and comprehensive perspective for the diagnosis of breast cancer.Thus, methods based on Graph Neural Networks (GNN) \cite{35kipf2016semi} have been increasingly used in breast cancer tissue classification \cite{10zhou2019cgc,11pati2022hierarchical,12patel2023garl}. These methods effectively handle the diversity and complexity of tissues in pathological images by capturing complex relationships between cells and tissues and providing deep feature representations. On the other hand, self-supervised learning methods \cite{31li2021dual,36li2023self,37koohbanani2021self,38srinidhi2022self,39ciga2022self} utilize large amounts of unlabeled data to learn complex patterns and structures within breast tissue images, thereby enhancing feature extraction, improving the generalization capability of the model, and reducing reliance on costly annotated data.

\subsection{Feature aggregation}

Feature aggregation plays a crucial role in machine learning and computer vision, especially in applications that require extracting meaningful information from multiple features or multiple data sources \cite{40ni2023feature,41zhang2020feature,42mercan2019patch,43touvron2021augmenting}. These methods aim to merge multiple input features into a comprehensive representation for effective information processing and decision-making.

Simpler forms of feature aggregation include average pooling and max pooling. Average pooling preserves background information by computing the mean of all features, while max pooling emphasizes the most significant features in an image by selecting the most prominent textures or patterns. Although these methods are straightforward to implement, they may not be sufficiently effective when dealing with complex or noisy data as they do not differentiate between the importance of features, potentially leading to the loss of critical information. To overcome these limitations, more advanced feature aggregation methods such as Weighted Average Pooling and Attention-based Aggregation have been developed. These methods assign different weights to different features, enabling more precise capture of key information in images or data. For instance, the Agg-Penultimate method \cite{42mercan2019patch} utilizes a weighted aggregation of the penultimate layer activations of the VGG16 network with class-specific Softmax outputs, providing an effective way to aggregate complex features from regions of interest (ROI). This method enhances the classification accuracy of ROIs and boosts the model's ability to distinguish between features of different categories. \cite{44cao2021feature} introduces a novel capsule network architecture that utilizes dual attention mechanisms to enhance feature aggregation for robust object tracking. Meanwhile, \cite{43touvron2021augmenting} replaces the final average pooling with an attention-based aggregation layer akin to a single transformer block, which weights how patches contribute to the classification decision. Additionally, \cite{36li2023self} proposes a global–local feature fusion prediction network based on the attention mechanism, designed to improve the survival prediction effect of WSIs with a comprehensive multi-scale information representation. These advanced techniques demonstrate significant improvements in handling and utilizing complex data effectively, which is critical in fields such as medical imaging and autonomous systems.

\section{Method}
\subsection{Overview}
\begin{figure*}
	\centering
		\includegraphics[scale=.5]{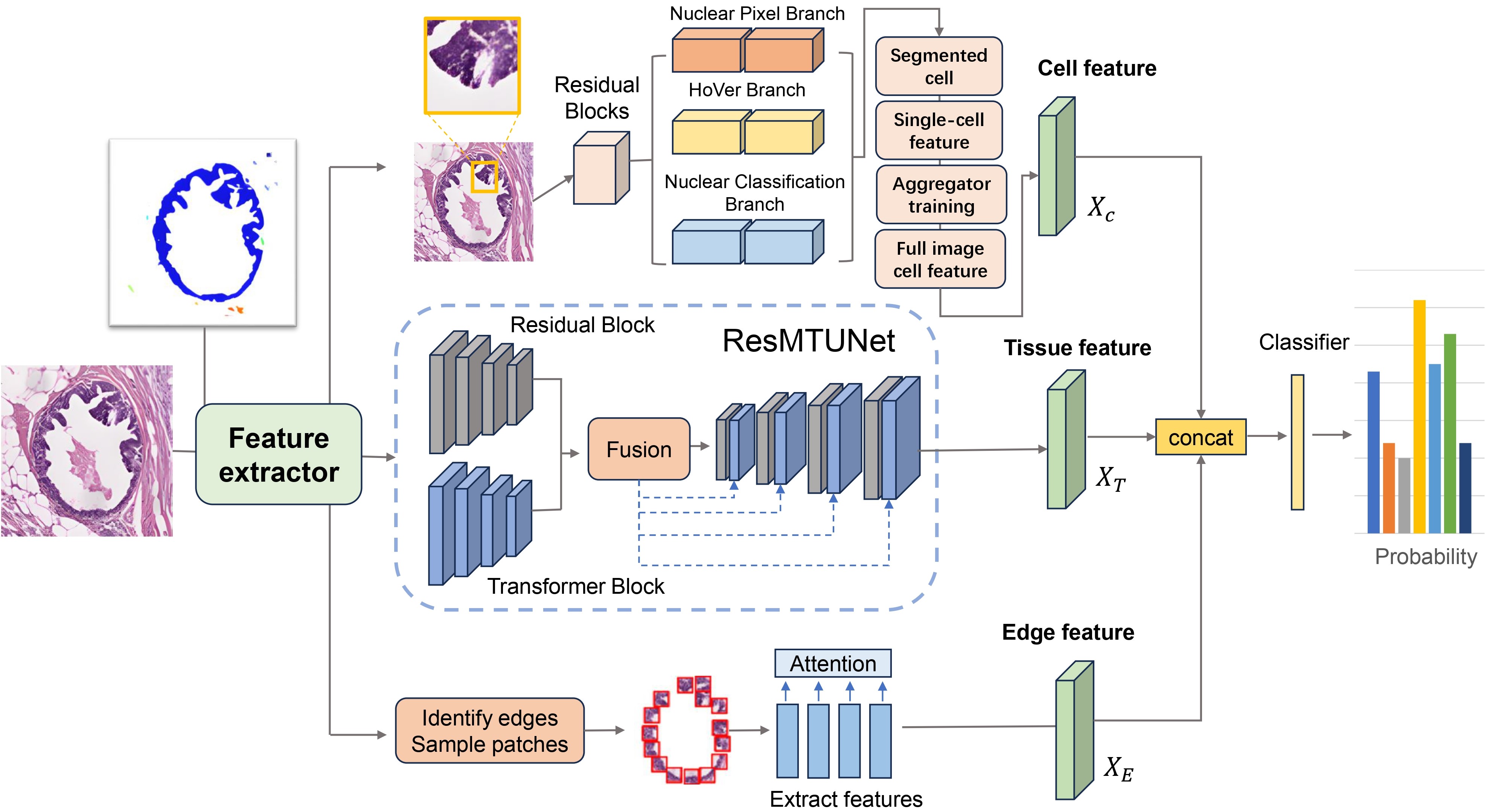}
	\caption{Breast cancer tissue image classification framework based on fusion features.}
	\label{FIG:1}
\end{figure*}
Figure \protect\ref{FIG:1} presents the architecture of a feature fusion method for breast cancer tissue image classification. This framework comprises key modules such as a cell feature extractor, tissue feature extractor, edge feature extractor, and a fusion feature classifier.

The architecture's utilization of multiple feature extractors to independently extract different features rather than relying on a single, unified feature extractor is primarily driven by several considerations. Firstly, this approach provides greater flexibility, allowing each extractor to be optimized for specific types of data or tasks. Such independence enables each extractor to be improved separately from the others, thereby enhancing the overall adaptability and generalization capability of the model.  Secondly, the use of multiple feature extractors helps to decouple the feature extraction process, breaking it down into smaller, more manageable parts. This decomposition increases the model's interpretability, as the output of each extractor can be analyzed and understood independently. When the model underperforms, it allows for more precise identification of the problem areas, facilitating targeted optimizations. Lastly, employing multiple feature extractors can also help reduce overfitting. A single extractor may lead to a model overly relying on certain specific features, thus increasing the risk of overfitting. Each part of the framework will be introduced in the following sections to provide a deeper understanding of its functionality and advantages.

\subsection{Cell feature extraction}
Cell features are critical for pathologists in the classification of breast tissue images. For instance, the presence or absence of myoepithelial cells is used to distinguish between in situ and invasive carcinoma, while cytological atypia helps in determining the presence and degree of lesions. Our study utilizes a multi-stage cell feature extractor to derive a comprehensive cell feature representation from breast epithelial tissue images. The process of whole-cell feature extraction can be broadly categorized into the following stages:

\textbf{Foreground tissue and cell segmentation:} Let $I \in \mathbb{R}^{H \times W}$ represent the input histological image, where $H$ and $W$ denote the image's height and width, respectively. Applying the cell segmentation model $S(\cdot)$, we obtain the epithelial tissue region mask $M$, where $M = S(I)$ and $M \in \{0,1\}^{H \times W}$, with $1$ indicating the epithelial tissue areas and $0$ indicating non-epithelial areas. Cell segmentation is performed using the HoverNet model \cite{45graham2019hover}. HoverNet is a deep learning network capable of simultaneous nucleus segmentation and classification, illustrated at the top branch in Figure \ref{FIG:1}. This network leverages rich instance information encoded within the vertical and horizontal distances from nucleus pixels to their centroids. These distances are then utilized to separate clustered nuclei, enabling precise segmentation, especially in areas with overlapping nuclei.

\textbf{Single cell feature extractor training:} Define the single cell feature extractor as a function $f_{\text{cell}}: \mathbb{R}^{h \times w} \rightarrow \mathbb{R}^d$, where $d$ represents the dimensionality of the feature space, and $h$ and $w$ are the dimensions of the local window size around a cell region. The feature extractor is trained on each cell location $x_i \in X$, where $X$ represents the set of all coordinates in the image, and $x_i$ is located within the epithelial tissue mask region $M$. The training process can be framed as an optimization problem:

\begin{equation}
\min_{f_{\text{cell}}} \sum_{i} L\left(f_{\text{cls}}\left(f_{\text{cell}}\left(I_{x_i}\right)\right), y_i\right)
\end{equation}

Where $L$ is the loss function, $y$ is the class label corresponding to the image. $f_{\text{cls}}$ is a temporary classifier, consisting of a fully connected layer, used to convert the features obtained by the feature extractor into class probability predictions.

\textbf{Mask area feature extraction:} Using the trained feature extractor $f_{\mathrm{cell}}$, extract features for each cell within the masked area: $F_i = f_{\mathrm{cell}}\left(I_{x_i}\right)$, where $F_i$ is the feature vector of cell $i$.

\textbf{Aggregator training:} Define the aggregator as the function $g: \mathbb{R}^{n \times d} \rightarrow \mathbb{R}^D$, where $n$ is the number of cells within the masked area, and $D$ is the dimension of the feature space after aggregation. The aggregator extracts the cell feature representation of the entire image by integrating the features of individual cells: $\mathbf{F}_{\mathrm{global}} = g(\{\mathbf{F}_i\}_{i=1}^n)$, where $\mathbf{F}_{\mathrm{global}} \in \mathbb{R}^D$ is the feature representation of the entire image. Similar to the training of the single-cell feature extractor, the optimization objective for training the aggregator is:
\begin{equation}
\min_g \sum_{i} L\left(f_{\text{cls}}\left(g(\{\mathbf{F}_i\}_{i=1}^n)\right), y\right) \quad
\end{equation}

\textbf{Whole image feature extraction:} Ultimately, use the trained aggregator $g$ to extract the cell-level features of the entire image, facilitating subsequent analysis or classification tasks.

\subsection{Tissue feature extraction}
Complementary to the cell feature extractor, the tissue feature extractor focuses on the overall tissue structure within the image. Let the tissue structural feature representation in the image be denoted as $X_t$, which is obtained through preprocessing and feature extraction operations as $X_t = f(I)$, where $f(\cdot)$ represents the feature extraction function. In this study, ResMTUNet is chosen as the backbone network for the feature extraction model. ResMTUNet is a novel architecture based on an encoder-decoder framework, combining the capabilities of modeling global features using ViT and local features using CNN, utilizing a UNet\cite{46ronneberger2015u} structured decoder to merge these multi-level features to generate aggregated feature maps for semantic segmentation, which can effectively distinguish the invasive cancer area from the ductal carcinoma in situ in whole slide images\cite{19liu2024invasive}. Similarly, the tissue feature extractor will be trained using the images from the training set and their class labels.

\subsection{Edge feature extraction}
\begin{figure*}
	\centering
		\includegraphics[scale=.55]{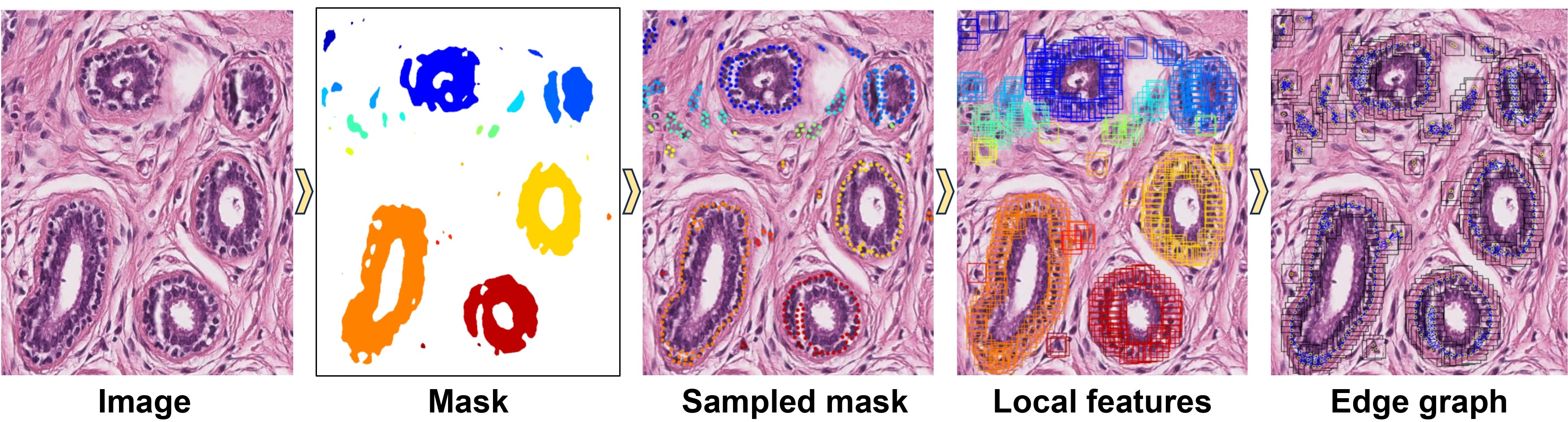}
	\caption{Edge feature extraction workflow.}
	\label{FIG:2}
\end{figure*}

From the cases we have collected, it can be observed that the edges of cancerous tissues are not the typical cord-like structures, making it difficult to discriminate based solely on conventional morphological features. More importantly, there is a subtle peripheral structure around the cancerous tissues in these cases, and even experienced pathologists need to use immunohistochemical methods such as staining with P63 to stain the myoepithelial cells in order to make a more accurate judgment. In fact, edge features play a crucial role in accurately distinguishing invasive cancer from in situ cancer. Therefore, to enhance the accuracy of classification, we propose an innovative graph-based edge feature extraction and fusion method, the process of which is shown in Figure \ref{FIG:2}.

We first divide the mask into multiple regions through connected-component analysis and generate a binary mask for each region. Next, we extract the contours of these masks and sample uniformly on the contours to obtain edge sampling points of the mask, reducing the computational load. Then, centered on these sampling points, we crop out $64\times64$ size patches. These patches, extracted at the edges, are preprocessed and then trained using a Vision Transformer model pretrained on the ImageNet dataset\cite{47deng2009imagenet} to obtain their feature representations. We integrate these features using a Nyström Attention\cite{48xiong2021nystromformer} aggregator trained to handle high-dimensional features such as edge features effectively, capturing the global dependencies among the features. Finally, using the KNN algorithm, we construct a method based on the image and sampling points to generate a nuclei map and combine features with sampling point data to construct an edge graph. This graph, composed of nodes and edges, where nodes represent cells or sampling points in the image and edges represent the relationships based on the features of the nodes and their spatial relations. This edge graph combines spatial information with the visual features of the cells, aiding in identifying which cell features are associated with their spatial positions, potentially revealing specific patterns in the tissue, which benefits the accuracy of classification.

\subsection{Multimodal feature fusion}
This stage involves the fusion of cell, tissue, and edge features to enhance the comprehensive evaluation of image content. The fused features are then input into a Support Vector Machine (SVM) classifier for classification, thereby achieving automated classification and assessment of the images. The Support Vector Machine is a supervised learning algorithm whose goal is to find an optimal hyperplane that effectively separates data points of different classes. Specifically, SVM determines the best decision boundary by maximizing the margin to achieve classification of the data.

Let the tissue feature be represented as $X_T$, the cell feature as $X_C$, and the edge feature as $X_E$. We adopt a weighted fusion strategy, where the weights illustrate the influence of each feature on the decision-making process. Higher weights mean that the corresponding feature plays a more critical role in the model. By appropriately assigning feature weights, the impact of information redundancy and noise can be effectively reduced, thereby enhancing the model's prediction accuracy and generalization performance. The fused feature representation is given by $X_f = [\alpha X_C, \beta X_T, \gamma X_E]$, where $[\cdot]$ denotes the concatenation operation, and $\alpha$, $\beta$, $\gamma$ respectively represent the weights for the cell, tissue, and edge features. The fused features $X_f$ are then input into the SVM classifier, whose objective function can be represented as follows:
\begin{equation}
\begin{split}
\min_{\mathbf{w},b,\xi_i} \frac{1}{2} \|\mathbf{w}\|^2 + C \sum_{i=1}^n \xi_i \\
\text{s.t.} \quad y_i(\mathbf{w} \cdot x_i + b) \geq 1 - \xi_i, & \quad \xi_i \geq 0
\end{split}
\end{equation}

Where $\mathbf{w}$ is the normal vector of the hyperplane, $b$ is the bias term, $C$ is the regularization parameter, $x_i$ is the feature vector of the $i$th sample, $y_i$ is the class label of the $i$th sample, $\xi_i$ is the slack variable, and $n$ is the number of samples. By using the SVM classifier trained on the fused features $X_f$, we can evaluate and classify new image data.

\subsection{Dataset}

To demonstrate the efficacy of our method convincingly, we opted to conduct experiments on the large-scale, publicly available BRACS dataset \cite{20brancati2022bracs}. BRACS provides an extensive collection of annotated Hematoxylin and Eosin (H\&E) stained images, including 547 whole slide images (WSIs) and 4539 regions of interest (ROIs) extracted from these WSIs. The ROIs were annotated using QuPath\cite{49bankhead2017qupath}, labeled as Normal (N), Pathological Benign (PB), Usual Ductal Hyperplasia (UDH), Atypical Ductal Hyperplasia (ADH), Flat Epithelial Atypia (FEA), Ductal Carcinoma In Situ (DCIS), and Invasive Carcinoma (IC). Our task is to develop a novel algorithm capable of achieving high classification accuracy across these seven categories in the BRACS dataset.

\begin{table*}[ht]
\centering
\caption{\textbf{Comparison with the prior art for breast cancer tissue image classification on the BRACS test dataset}, including the F1 score for each category and the weighted F1 score for seven-category classification. The results are presented in percentages(\%). The best results are highlighted in bold, and the second-best results are underlined.}
\label{tab1}
\begin{tabular}{lcccccccc}
\toprule
Model & N & PB & UDH & ADH & FEA & DCIS & IC & Weighted F1 \\
\midrule
Patch-GNN \cite{51aygunecs2020graph} & $52.5 \pm 3.3$ & $47.6 \pm 2.2$ & $23.7 \pm 4.6$ & $30.7 \pm 1.8$ & $60.7 \pm 5.3$ & $58.8 \pm 1.1$ & $81.6 \pm 2.2$ & $52.1 \pm 0.6$ \\
CGC-Net \cite{10zhou2019cgc} & $30.8 \pm 5.3$ & $31.6 \pm 4.7$ & $17.3 \pm 3.4$ & $24.5 \pm 5.2$ & $59.0 \pm 3.6$ & $49.4 \pm 3.4$ & $75.3 \pm 3.2$ & $43.6 \pm 0.5$ \\
CG-GNN \cite{11pati2022hierarchical} & $63.6 \pm 4.9$ & $47.7 \pm 2.9$ & $39.4 \pm 4.7$ & $28.5 \pm 4.3$ & $72.1 \pm 1.3$ & $54.6 \pm 2.2$ & $82.2 \pm 4.0$ & $56.6 \pm 1.3$ \\
TG-GNN \cite{11pati2022hierarchical} & $58.8 \pm 6.8$ & $40.9 \pm 3.0$ & $\underline{46.8 \pm 1.9}$ & $40.0 \pm 3.6$ & $63.7 \pm 10.5$ & $53.8 \pm 3.9$ & $81.1 \pm 3.3$ & $55.9 \pm 1.0$ \\
HACT-Net \cite{11pati2022hierarchical} & $61.6 \pm 2.1$ & $47.5 \pm 2.9$ & $43.6 \pm 1.9$ & $40.4 \pm 2.5$ & $74.2 \pm 1.4$ & $\underline{66.4 \pm 2.6}$ & $88.4 \pm 0.2$ & $61.5 \pm 0.9$ \\
CLAM \cite{32lu2021data} & $59.4 \pm 2.0$ & $47.7 \pm 1.2$ & $31.7 \pm 0.7$ & $20.1 \pm 3.4$ & $68.3 \pm 0.4$ & $59.9 \pm 1.7$ & $86.8 \pm 0.6$ & $54.8 \pm 1.0$ \\
TransMIL \cite{13shao2021transmil} & $47.6 \pm 9.8$ & $42.9 \pm 3.6$ & $41.5 \pm 5.3$ & $38.4 \pm 5.9$ & $72.7 \pm 2.6$ & $62.7 \pm 2.9$ & $87.1 \pm 3.9$ & $57.5 \pm 0.7$ \\
ScoreNet \cite{15stegmuller2023scorenet} & $\underline{64.3 \pm 1.5}$ & $\bm{54.0 \pm 2.2}$ & $45.3 \pm 3.4$ & $\bm{46.7 \pm 1.0}$ & $\underline{78.1 \pm 2.8}$ & $62.9 \pm 2.0$ & $\underline{91.0 \pm 1.4}$ & $\underline{64.4 \pm 0.9}$ \\
FECT (ours) & $\bm{75.5 \pm 2.6}$ & $\underline{51.8 \pm 1.3}$ & $\bm{47.0 \pm 1.7}$ & $\underline{45.2 \pm 2.2}$ & $\bm{79.6 \pm 2.6}$ & $\bm{66.7 \pm 2.4}$ & $\bm{94.3 \pm 1.3}$ & $\bm{65.8 \pm 0.8}$ \\
\bottomrule
\end{tabular}
\end{table*}

To enhance our classification task, we incorporated the seven-category segmentation masks for BRACS breast tissue, as proposed in \cite{19liu2024invasive}. This work outlines a semi-automated annotation process divided into pre-segmentation and fine-tuning phases. Initially, a small sample set is extracted from the BRACS training set, and epithelial tissue contours are annotated using QuPath. A lightweight UNext network \cite{50valanarasu2022unext} is then trained to model pre-segmentation, which is used to infer on a portion of the unannotated samples. Poorly segmented results are manually annotated. This cycle is repeated until the model performing best on the validation set is used to infer on all data and generate the final epithelial segmentation masks. This method produces a segmentation mask for each ROI image, guiding the classification model to better understand image edge contours and extract features from a morphological perspective.

\section{Experiments}
\subsection{Experimental setup and metrics}
When training the cell, tissue, and edge feature extractors, the ResMTUNet model and Nyström Attention aggregator will undergo iterative training over 30 epochs, with a batch size set at 16 for each epoch. The initial learning rate is set at 0.001, and a step decay strategy is applied, where the learning rate is halved every 7 training cycles. Additionally, the SGD algorithm with a momentum of 0.9 is employed for optimization to effectively adjust model weights, enhancing training efficiency and model performance. All images from the BRACS dataset are uniformly resized to a resolution of $512\times512$ pixels before being fed into the model to ensure the data input is consistent. 

To enhance the model's generalizability, a variety of data augmentation techniques are used, including horizontal and vertical flips, Gaussian noise, and a series of blurring and sharpening processes, such as motion blur and median blur. Additionally, random adjustments to image brightness and contrast are implemented to help maintain robustness and accuracy of the model. All computations for these experiments were conducted on single 24 GB NVIDIA 3090 GPUs using the PyTorch framework.

The classification of BRACS breast cancer tissue images is quantitatively evaluated using two different metrics: Overall classification accuracy (Acc) and the Arithmetic mean of the per-class F1-scores (F1). Accuracy is calculated by computing the unweighted average of recall for each category, which accounts for the class imbalance in the evaluation set. The weighted F1 score is computed by averaging the F1 scores (the harmonic mean of precision and recall) of each category, weighted according to the support of each class. This method ensures a more balanced evaluation across varying class sizes and complexities.

\subsection{Results}

We first demonstrate that our FECT model outperforms the current state-of-the-art methods for breast cancer classification. We compare the classification performance of FECT on the BRACS test set with various approaches including Graph Neural Networks (GNN-based methods such as Patch-GNN\cite{51aygunecs2020graph}, CGC-Net\cite{10zhou2019cgc}, TG-PNA\cite{11pati2022hierarchical}, CG-PNA\cite{11pati2022hierarchical}, HACT-Net\cite{11pati2022hierarchical}), Multiple Instance Learning (MIL-based method CLAM\cite{32lu2021data}), and Transformer-based methods (TransMIL\cite{13shao2021transmil}, ScoreNet\cite{15stegmuller2023scorenet}). Among these, CLAM utilizes the best performing model, CLAM-MB/B (40×), on the BRACS dataset. ScoreNet employs the ScoreNet/4/3 configuration, which leverages the four last [CLS] tokens of the scorer and the three last [CLS] tokens from the coarse attention mechanism (aggregation stage).

FECT achieves a new state-of-the-art weighted F1-score of 65.8\% on the BRACS Tissue Regions of Interest classification task, outperforming the third-best method, HACT-Net\cite{11pati2022hierarchical}, by a margin of 4.3\%, and surpassing the second-best method, ScoreNet\cite{15stegmuller2023scorenet}, by 1.4\% (Table \ref{tab1}). Our proposed method FECT demonstrates substantial performance improvements in the majority of categories as evidenced in Table \ref{tab1}. Particularly, in the classification of normal tissue (N), FECT achieves a remarkable F1-score of 75.5\%, significantly higher than other methods. Moreover, FECT exhibits exceptional performance in recognizing Atypical Ductal Hyperplasia (ADH) and Ductal Carcinoma In Situ (DCIS) with F1-scores of 45.2\% and 79.6\% respectively. In the classification of Invasive Carcinoma (IC), FECT leads with an F1-score of 94.3\%, surpassing the second-highest score of ScoreNet, which is 91.0\%. Overall, FECT reaches a weighted F1-score of 65.8\% across seven categories, outstripping all comparison methods. These results highlight the efficacy and potential of the fused feature architecture in the classification of breast epithelial tissue images.

\begin{table}[ht]
\centering
\caption{\textbf{Comparison of weighted F1 scores and accuracy results(\%) across different feature fusion strategies}, namely concat, weighted fusion and tensor fusion. The top-performing results are highlighted in bold.}
\label{tab2}
\begin{tabular}{lcc}
\toprule
Model                             & Weighted F1 & Acc \\
\midrule
ResNet + HACT-Net \cite{52yu2023two}            & $55.18$         & $55.68$ \\
ViT + HACT-Net (Concat) \cite{52yu2023two}       & $61.57$         & $61.82$ \\
ViT + HACT-Net (Weighted) \cite{52yu2023two}     & $60.86$         & $60.86$ \\
ViT + HACT-Net (Tensor) \cite{52yu2023two}       & $64.88$         & $64.86$ \\
FECT (ours)                       & $\bm{65.75}$    & $\bm{66.37}$ \\
\bottomrule
\end{tabular}
\end{table}

We also compared the performance of different feature fusion strategies in breast cancer classification, as presented in Table \ref{tab2}. The comparison includes feature fusion strategies based on CNNs, Vision Transformers (ViT), and the state-of-the-art method for breast tissue pathology image classification, HACT-Net\cite{52yu2023two}. Specifically, tensor fusion of features extracted by ViT and HACT-Net was implemented by constructing the Cartesian product of multidimensional data, explicitly modeling the multimodal interactions between features. Due to the high dimensionality of edge features in the FECT method, tensor fusion resulted in a substantial computational burden; hence, we opted for a weighted fusion strategy. Despite this, our approach achieved the best performance in both weighted F1 scores and accuracy, with a weighted F1 score of 65.75\% and an accuracy of 66.37\%. This highlights the advantages of FECT in leveraging specialized models to extract multi-level features from pathological and morphological perspectives.

\subsection{Ablation study}

We conducted a series of experiments to demonstrate the value of integrating cell, tissue, and edge-level features for the classification of breast cancer tissue images. Table \ref{tab3} presents the performance metrics of several ablation variants of the FECT model on the BRACS test dataset classification task. These variants utilized tissue features, cell features, edge features, and their combinations as inputs to the classifier, with performance assessed through accuracy and F1 scores. The experimental results reveal that when cell, tissue, and edge features are used individually as inputs, the classifier's accuracy was 55.87\%, 61.92\%, and 51.60\% respectively, with F1 scores of 55.68\%, 61.49\%, and 49.38\%. However, when fused features were employed, the classifier's performance significantly improved, achieving an accuracy of 66.37\% and an F1 score of 65.75\%. This indicates that fused features can effectively leverage the information from structural and cell features, enhancing the classifier's accuracy and generalization ability. Notably, although using edge features alone resulted in lower classification accuracy, their contribution to enhancing classification accuracy when fused with other features cannot be overlooked. This suggests that edge features, although not robust enough alone, can effectively complement other features, boosting the model’s ability to recognize areas with indistinct boundaries.

\begin{table}[ht]
\centering
\caption{Classifier performance with different features.}
\label{tab3}
\begin{tabular}{lcc}
\toprule
Classifier input     & Acc (\%) & F1 (\%) \\
\midrule
Cell feature           & $55.87$      & $55.68$ \\
Tissue feature         & $61.92$      & $61.49$ \\
Edge feature           & $51.60$      & $49.38$ \\
Cell + Tissue feature  & $63.17$      & $62.48$ \\
Cell + Edge feature    & $60.14$      & $59.68$ \\
Tissue + Edge feature  & $64.06$      & $63.08$ \\
Fusion feature         & $\bm{66.37}$ & $\bm{65.75}$ \\
\bottomrule
\end{tabular}
\end{table}

\begin{figure}
	\centering
		\includegraphics[scale=.65]{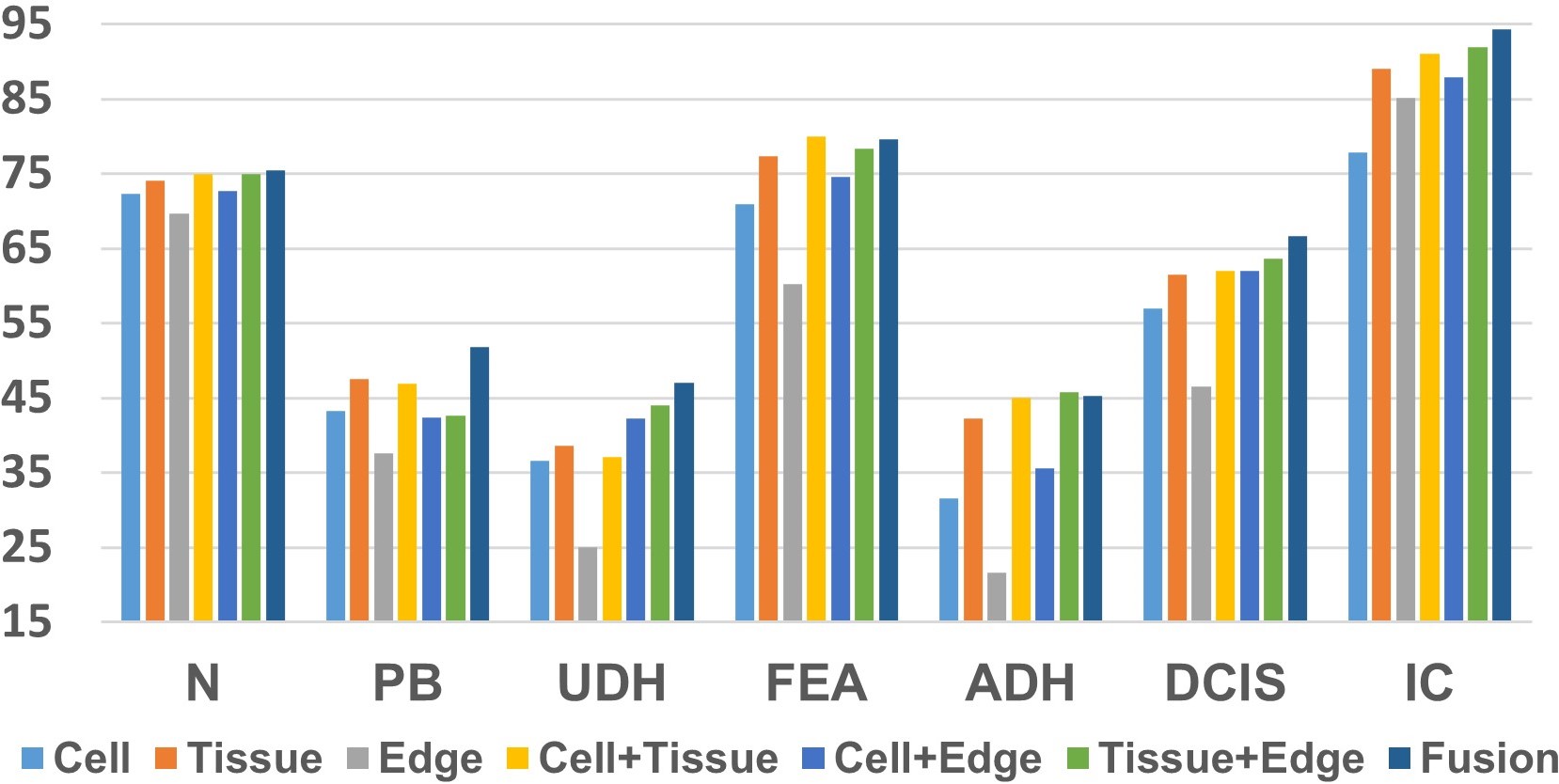}
	\caption{Comparison of F1 Scores (\%) for classifiers based on different features across seven categories of breast cancer tissues.}
	\label{FIG:3}
\end{figure}

In further analyzing the performance of classifiers, we conducted a detailed exploration using bar charts to compare the classification performance of tissue features, cell features, edge features, and fused features across seven categories. Figure \ref{FIG:3} visually displays the F1 scores of these three types of features and their combinations across different categories. The cell features showed high classification effectiveness in categories N and IC, with F1 scores reaching 72.29\% and 77.85\% respectively. However, in the categories PB, UDH, and ADH, the performance of cell features was not optimal, with F1 scores all below 50\%.In contrast, tissue features displayed better classification results in FEA and IC, with F1 scores of 77.27\% and 89.03\% respectively, but their performance in other categories needs improvement. Notably, fused features exhibited a trend of performance improvement in most categories, especially in categories such as PB, UDH, DCIS, and IC, significantly surpassing the performance of single features.

Additionally, upon discovering that some invasive carcinoma samples were still incorrectly classified as in situ carcinoma after fusing cell and tissue features, we considered integrating edge features to identify a subtle peripheral structure present around cancerous tissues. The experimental results demonstrated that after incorporating edge features, the classification accuracy for DCIS and IC improved, confirming that edge features, as a complementary feature, highlighted the contrast between tumor cells and normal cells, as well as between internal tumor structures and surrounding non-tumor tissues. This enhancement allowed the classification model to more easily recognize and learn these critical pathological differences, thus improving diagnostic accuracy.

The choice of classifier plays a crucial role in the performance of classification tasks. To validate the superior performance of our chosen SVM classifier, we conducted experiments with different classifiers using the same training and validation sets, comparing their performance metrics on the BRACS test set (Table \ref{tab4}). All machine learning classifiers were employed with default parameters, and the results demonstrated that the SVM classifier exhibited exceptional performance when handling breast cancer tissue images characterized by complex boundaries and high-dimensional feature spaces. Specifically, the SVM achieved higher accuracy and F1 scores compared to other classifiers such as Logistic Regression, Decision Trees, Random Forests, XGBoost, and LightGBM. Logistic Regression was faster in training but slightly lower in both accuracy (65.1\%) and F1 score (64.7\%). Decision Trees provided good interpretability but performed the worst across all metrics, likely due to a tendency to overfit the training data. Random Forests, XGBoost, and LightGBM also underperformed compared to SVM in this context.

\begin{table}[ht]
\centering
\caption{Comparison of classifier performance.}
\label{tab4}
\begin{tabular}{lcc}
\toprule
Classifier           & Acc (\%)     & F1 (\%) \\
\midrule
Logistic Regression  & $62.10$      & $61.35$ \\
Random Forest        & $61.21$      & $59.95$ \\
Decision Tree        & $56.41$      & $56.40$ \\
XGBoost              & $61.57$      & $60.79$ \\
LightGBM             & $61.21$      & $60.28$ \\
SVM                  & $\bm{66.37}$ & $\bm{65.75}$ \\
\bottomrule
\end{tabular}
\end{table}

\begin{figure}
	\centering
		\includegraphics[scale=.4]{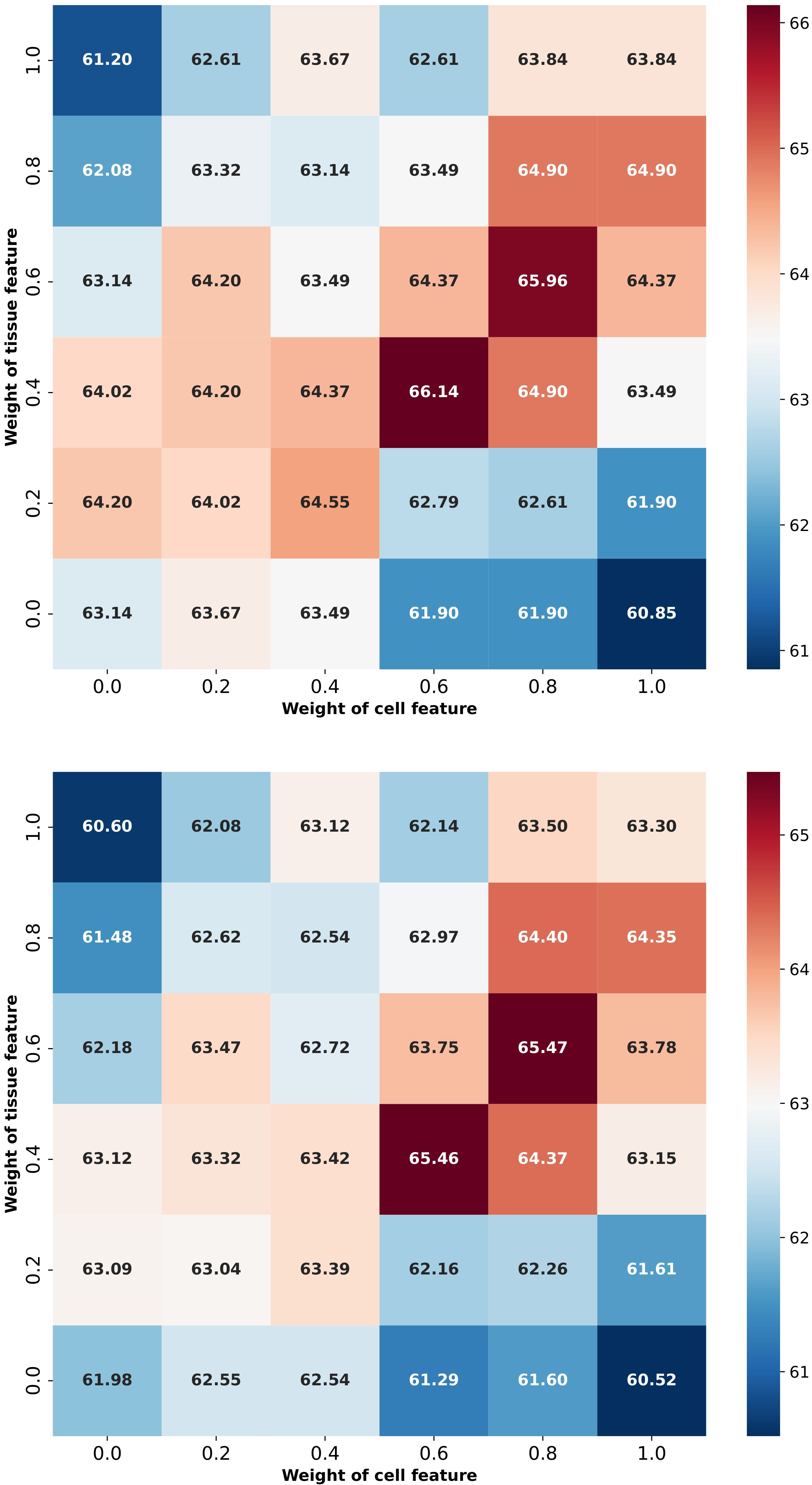}
	\caption{\textbf{Impact of different weight combinations of cell and tissue features on classification accuracy and F1 score.} The upper heatmap represents accuracy, while the lower one shows the F1 score (\%).}
	\label{FIG:4}
\end{figure}

During the feature fusion process, we adopted a weighted fusion approach, meaning that the weight assigned to each feature significantly impacts classification performance. Since edge features serve as complements to cell and tissue features, our study focused specifically on the impacts of cell and tissue feature weights. We employed grid search techniques to identify the optimal weight parameter combinations and visualized the impact of each parameter combination on the model's classification performance using heatmaps. As illustrated in Figure \ref{FIG:4}, the heatmaps reveal that the highest accuracy and F1 scores are found in a central region skewed slightly to the right, around a cell feature weight of approximately 0.6 and a tissue feature weight of approximately 0.4. In this region, both accuracy and F1 score exceed 65\%. When the weight of cell features approaches 1, regardless of the tissue feature weight, the accuracy remains relatively low, indicating that relying solely on cell features is insufficient for achieving high accuracy. Similarly, when the weight of tissue features is very low (near 0), even with a high weight for cell features, the accuracy is also relatively low, underscoring the indispensable role of tissue features in the classification process. 

\subsection{Visualization and interpretation}

\begin{figure}
	\centering
		\includegraphics[scale=.4]{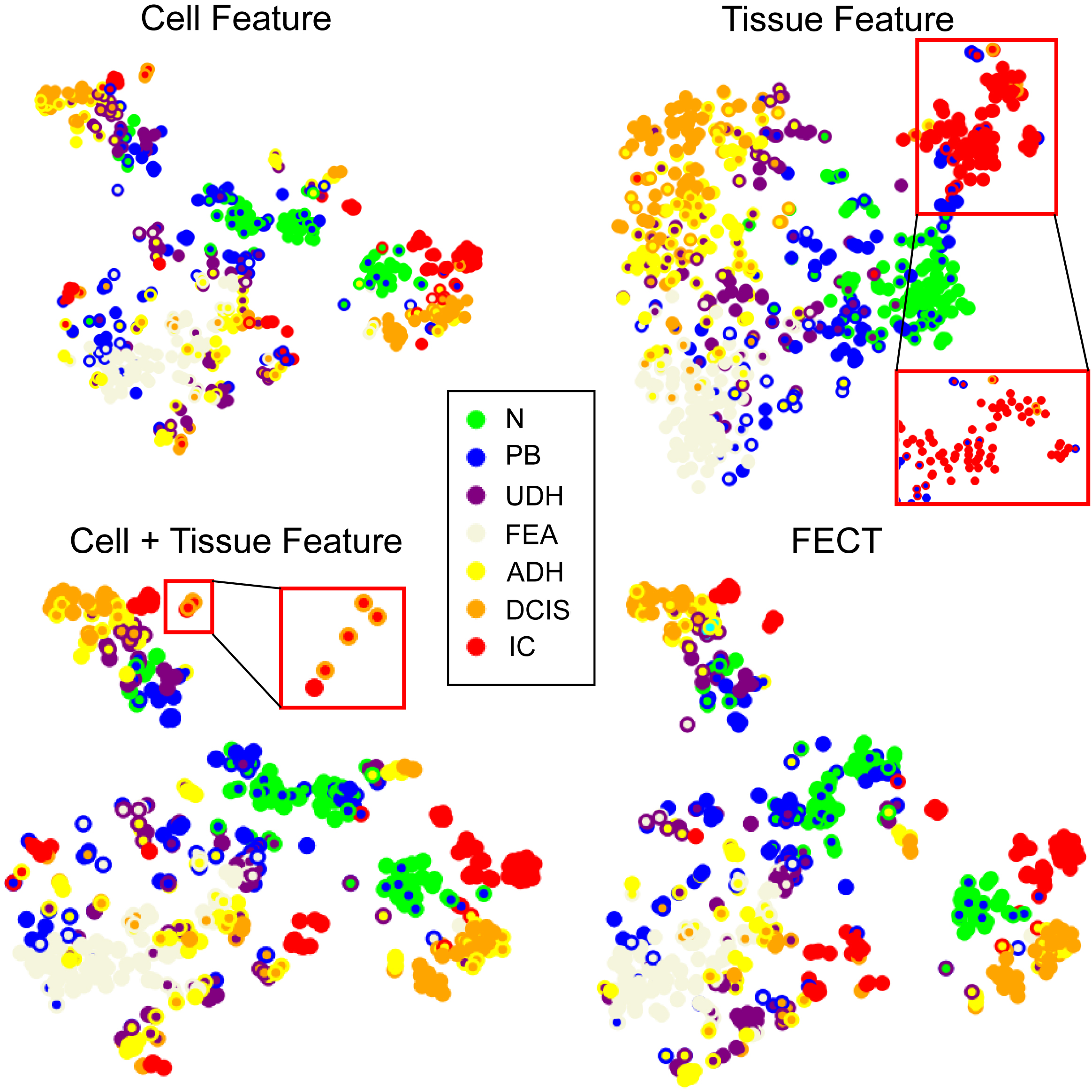}
	\caption{\textbf{The t-SNE dimensionality reduction visualization of cell, tissue, and edge features.} Each small dot represents a sample. The fill color of the dot indicates the actual category of the sample, while the outline color of the dot indicates the category to which it has been predicted.}
	\label{FIG:5}
\end{figure}

Our study employed t-SNE dimensionality reduction techniques to conduct an in-depth visual analysis of tissue features, cell features, edge features, and their fused features. Figure \ref{FIG:5} reveals the spatial layout of these three types of features after dimensionality reduction. From the dimensionality reduction map of tissue features, we can clearly see the clustering of different categories, although their spatial distribution appears somewhat loose. This reflects the effectiveness of the tissue feature extractor in capturing low-frequency features, which provide strong support for distinguishing different categories at the tissue level. However, its ability to differentiate between samples of different categories that are similar in tissue features is somewhat lacking. On the other hand, the dimensionality reduction map for cell features displays numerous tight and well-defined clusters, with the separation of certain category clusters being particularly notable. This reveals significant heterogeneity within the cell features of these categories, indicating that the cell feature extractor excels at capturing high-frequency features that can reveal subtle differences between samples. These high-frequency features, rich in detail, are crucial for identifying samples that are structurally similar but differ in cell characteristics. This capability is essential for accurately classifying samples where tissue features may overlap but cell features provide the necessary differentiation.

In Figure \ref{FIG:5}, we can observe the intra-class dispersion and inter-class overlap in different dimensionality reduction plots, reflecting the degree of differentiation between categories. For example, FEA exhibits low intra-class dispersion and minimal overlap with other categories across all four plots, indicating that it is more easily distinguishable. In contrast, normal categories (N, PB) and Invasive Carcinoma (IC) show a clear separation trend, demonstrating significant differences in feature space. Conversely, ADH shows high overlap with UDH and DCIS, indicating greater difficulty in differentiation.

\begin{figure}
	\centering
		\includegraphics[scale=.55]{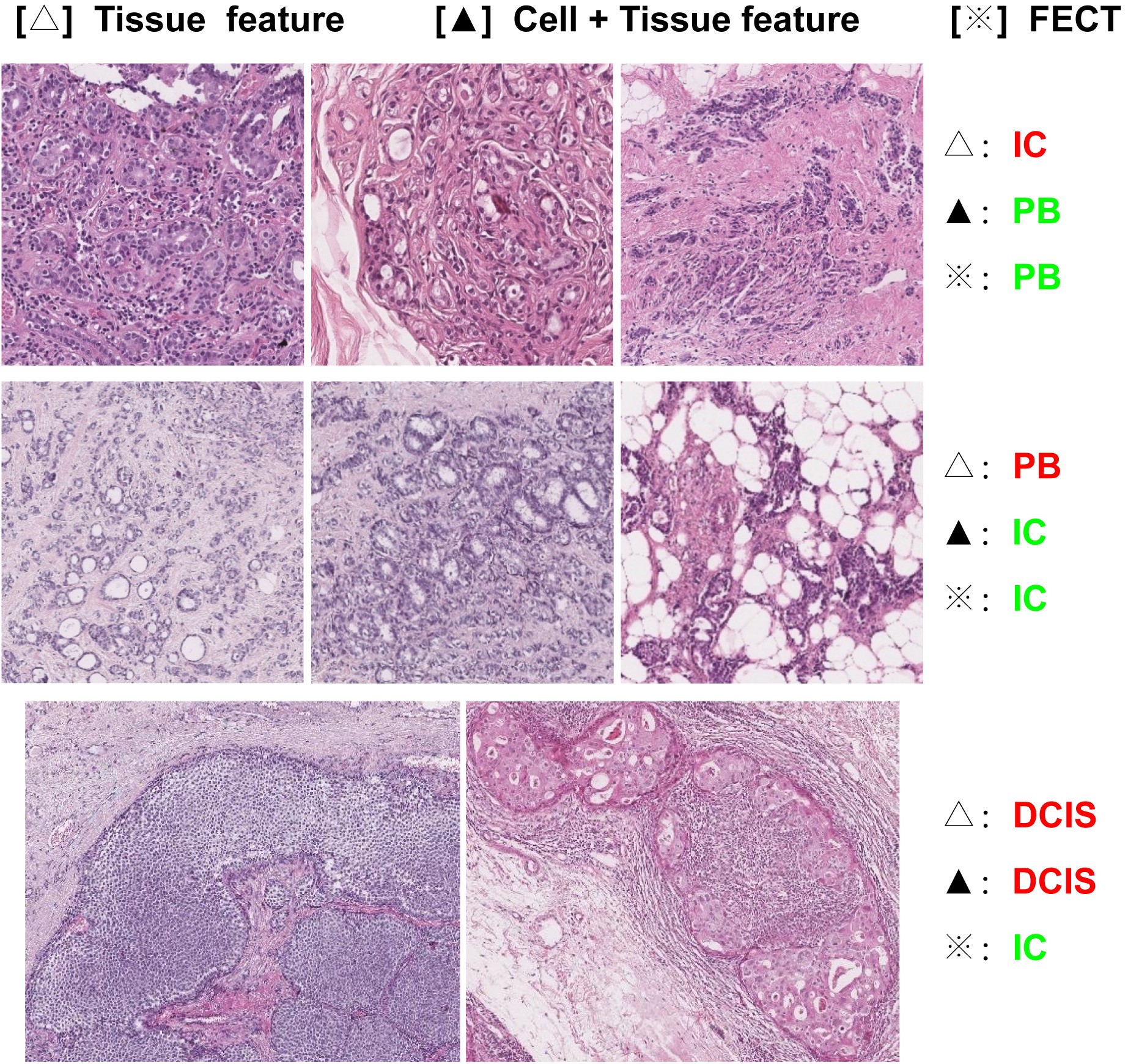}
	\caption{\textbf{Qualitative comparison of the classification effects of tissue features, cell-tissue features and cell-tissue-edge features on difficult-to-classify categories.} The classifier predictions are noted to the right of each example. Red and green indicate incorrect and correct classifications, respectively. If in the tissue feature classifier, in the first row, benign lesions are mistakenly classified as invasive cancer; in the second row, invasive cancer is mistakenly classified as benign lesions. These images were correctly classified in the cell-tissue fusion feature classifier. The two cases of invasive cancer in the third row were misclassified as carcinoma in situ even in the cell-tissue classifier, but were correctly classified in FECT.}
	\label{FIG:6}
\end{figure}

Notably, from the tissue feature dimensionality reduction plot (top right of Figure \ref{FIG:5}), we can clearly see that a considerable portion of benign lesions (PB) are incorrectly classified as invasive carcinoma (IC), and several invasive carcinomas are misclassified as benign lesions. However, the introduction of edge features significantly alleviates this classification confusion in the fused features. To further visually demonstrate these classification confusions, we have selected some typical cases depicted in Figure \ref{FIG:6}. It is evident that these samples exhibit high structural similarity, which is the primary reason why the tissue feature classifier tends to confuse them. However, the classifier utilizing both cell and tissue fusion features successfully corrects these misclassifications because cell features can help identify cell-level biomarkers, such as nuclear size, morphology, and chromatin patterns, all critical factors in pathological diagnosis.

Despite the classifier using cell and tissue fusion features, there are still some cases where invasive carcinoma samples are incorrectly classified as in situ carcinoma. In classifiers that fuse all three types of features, these samples are correctly classified, due to the subtle peripheral structure of the cancerous tissues in these cases, making the differences in edge features more significant. Specifically, edge features include information about the shape, texture, or relative positioning of lesion boundaries to surrounding tissues—details not captured by cell or tissue features alone. These provide a more accurate basis for classification, supporting the model's interpretability and clinical applicability.

\section{Discussion and conclusion}
The primary challenges in automatically classifying breast cancer subtypes from digital pathology images include high heterogeneity, complex microstructures, and the need for models that offer both explainability and accuracy. In this paper, we introduce a model that applies Fused features of Edges, Cells, and Tissue (FECT) to automatically classify breast cancer subtypes from digital pathology images. We demonstrate the superior classification performance of our model on the BRACS test set, achieving state-of-the-art results that surpass other models. This contributes to enhancing the accuracy and efficiency of breast cancer pathological diagnoses in clinical settings.

Previous methods for classifying breast tissue images have laid the foundation for developments in this field. Table \ref{tab1} presents a quantitative comparison of our method with past approaches. \cite{51aygunecs2020graph} proposed a generic methodology that incorporates local inter-patch context through a graph convolution network (GCN), achieving a weighted F1 score of 52.1\%. \cite{11pati2022hierarchical} introduced HACT-Net, which utilizes well-defined cells and tissue regions to build Hierarchical Cell-to-Tissue (HACT) graph representations, reaching a weighted F1 score of 61.5\%. \cite{32lu2021data} presented CLAM, which uses attention-based learning to identify subregions of high diagnostic value to accurately classify whole slides, achieving a weighted F1 score of 54.8\%.  \cite{13shao2021transmil} proposed TransMIL, a Transformer-based MIL that achieved a weighted F1 score of 57.5\%. \cite{15stegmuller2023scorenet} developed an efficient new transformer that exploits a differentiable recommendation stage to extract discriminative image regions, demonstrating high performance with a weighted F1 score of 64.4\%. Finally, \cite{52yu2023two} designed a weighted tensor feature fusion module to fuse the information extracted from ViT and GNN, which achieved a weighted F1 score of 64.88\% and an accuracy of 64.86\%. These advancements illustrate significant progress in the field, with our model setting a new benchmark for future research and clinical applications.

In our method, we achieved a weighted F1 score of 65.75\% and an accuracy rate of 66.37\%. Experiments demonstrate that our approach, which fuses features of cells, tissues, and edges, can significantly enhance the accuracy and reliability of breast cancer tissue image classification. This improvement is attributable to two main factors: Firstly, the independence of each feature extractor allows for individual improvements independent of others, enhancing model flexibility, adaptability, and generalizability. Secondly, our model is designed with clinical interpretability in mind, training features based on the morphological characteristics of specific pathological image categories and exploring spatial relationships within the images, aligning with the diagnostic processes of pathologists and making it easy to understand.

Despite the promising results demonstrated by the proposed FECT, there are still some limitations. Firstly, there is room to improve the model's classification accuracy. As seen in Table \ref{tab1} and Figure \ref{FIG:5}, UDH  and ADH remain challenging to classify and are prone to mutual confusion. This is partly due to their very similar cell and tissue structures in pathological images, which even experienced pathologists can find difficult to distinguish. Future work could focus on designing specific features based on their morphological characteristics to be extracted and integrated into the classifier. Secondly, while the edge feature extractor in FECT acts as a complement to the cell and tissue feature extractors, it performs poorly when used alone as a classifier. Consideration could be given to improving the way edge features are extracted, possibly using a more adaptive and generalized attention model to train the patch aggregator. Another approach could be to enhance feature fusion methods, such as tensor fusion, which can capture correlations between different modal features, though this is highly memory-intensive.

In future work, we plan to continue our research in two main directions. First, the current model primarily relies on WSI. Future studies could consider integrating various types of data, such as genetic data and clinical indicators, to enhance the predictive performance and robustness of the model. Multimodal feature fusion could reveal associations between complex biomarkers that image data alone cannot provide, offering a more comprehensive approach to breast cancer typing and diagnosis. Second, as the current model has only been trained and tested on the publicly available BRACS dataset, to enhance the practical value of our model, close collaboration with clinical physicians is essential to drive clinical applications. Through such collaborations and feedback, further learning can be promoted to ensure high diagnostic accuracy across various clinical settings.

\section*{Funding}
The work was supported by the Shenzhen Engineering Research Centre (XMHT20230115004), the Shenzhen Science and Technology Innovation Commission (KCXFZ2020\\1221173207022), and the Jilin Fuyuan Guan Food Group Co., Ltd.

\section*{Data availability statement}
You can access the BRACS dataset through the provided link: \href{https://www.bracs.icar.cnr.it/}{https://www.bracs.icar.cnr.it/}.

\section*{Declaration of competing interest}
The authors declare that they have no known competing financial interests or personal relationships that could have appeared to influence the work reported in this paper.

\printcredits

\bibliographystyle{model1-num-names}
\bibliography{cas-refs}

\end{document}